# Dynamics of the magnetic and structural α–ε phase transition in Iron

O. Mathon[1], F. Baudelet[2,3], J.P. Itié[2], A. Polian[2], M. d'Astuto[2], J.C. Chervin[2] and S. Pascarelli[1]

[1] *European Synchrotron Radiation Facility, BP 220, 38043 Grenoble CEDEX, France.*
[2] *Physique des Milieux Condensés (UMR 7602), Université Pierre et Marie Curie B77, 4 Place Jussieu, 75252 Paris CEDEX 05, France.*
[3] *Synchrotron SOLEIL, L'Orme des Merisiers, Saint-Aubin, BP 48, 91192 Gif-sur-Yvette CEDEX, France.*

Abstract:

We have studied the high-pressure iron bcc to hcp phase transition by simultaneous X-ray Magnetic Circular Dichroism (XMCD) and X-ray Absorption Spectroscopy (XAS) with an X-ray dispersive spectrometer. The combination of the two techniques allows us to obtain simultaneously information on both the structure and the magnetic state of Iron under pressure. The magnetic and structural transitions simultaneously observed are sharp. Both are of first order in agreement with theoretical prediction. The pressure domain of the transition observed (2.4 ± 0.2 GPa) is narrower than that usually cited in the literature (8 GPa). Our data indicate that the magnetic transition slightly precedes the structural one, suggesting that the origin of the instability of the bcc phase in iron with increasing pressure is to be attributed to the effect of pressure on magnetism as predicted by spin-polarized full potential total energy calculations.

The iron phase-diagram has attracted considerable interest since a long time. At the beginning the motivation was its central role in the behaviour of alloys and steel [1]. Later its geophysical importance was underlined, because of its predominant abundance in the Earth's core [2] [3]. The phase diagram of iron at extreme pressure and temperature conditions is still far from being established. Under the application of an external pressure, iron undergoes a transition at 13 GPa from the bcc α-phase to the hcp ε-phase structure [4], with the loss of its ferromagnetic long range order [5]. In the literature the transition extends over 8 GPa at ambient temperature within which the bcc and hcp phases coexist [6]. In the pure hcp ε-phase, superconductivity appears close to the transition between 15 and 30 GPa [7,8].

The evolution with pressure of the magnetic and structural state across this transition is still a subject of active theoretical and experimental research.

Considerable *ab initio* theoretical studies of this transition have been carried out[1] [2],[8],[9],[10], [11], [12] [13]. The first order nature of the structural and magnetic transitions is well established as well as the non-ferromagnetic state of the hcp phase. It is also clear that ferromagnetism of Iron has a fundamental role for the ground state structure in the stabilization of the bcc phase with respect to hcp or fcc. Ref. [9] for the first time predicts the driving role of magnetism in the α–ε phase transition. More recently, Ref. [10] presents a complete map of the transition path between bcc and hcp phase using a spin-polarized full-potential total energy calculation within the generalized-gradient method (GGA). It is found that the effect of pressure is essentially to broaden the d bands and to decrease the density of states at the Fermi level below the stability limit for ferromagnetism given by the Stoner criterion. Therefore, they conclude that the α–ε phase transition is not primarily due to phonon softening but to the effect of pressure on the magnetism of Iron. Recent calculations [14] based on a multiscale model containing a quantum-mechanics based multiwell energy function proposes an initiation of the bcc-hcp phase transition primarily due to shear. During the transition a low spin state is also suggested. At higher densities, an incommensurate spin-density-wave-ordered state is predicted in Ref. 15 for a small pressure range starting with the onset of the hcp phase, as for superconductivity.

Structural investigations have been reported using X-ray diffraction [16] and more recently EXAFS [6]. The iron bcc-hcp phase transition is described as a martensitic transition with a slow variation of the relative bcc and hcp phases abundance. Ref. 6 reports distorted bcc and hcp phases with anomalously large lattice constant (bcc) and c/a ratio (hcp) when the relative amounts become small. These anomalies are attributed to interfacial strain between the bcc and the hcp phases. A transition model based on lattice shearing movements is proposed with a possible intermediate fcc structure.

Magnetic measurements are traditionally based on Mössbauer techniques [5,17,18,19] and, recently, on inelastic x-ray scattering [20]. The authors of Ref. 20 report on variation of a satellite in the Fe-K$\beta$ fluorescence line, which shows that the magnetic moment of iron decreases over the same pressure range as the bcc to hcp transition i.e. 8 GPa. These data agree with Mössbauer results [19], which indicate that the main part of the magnetic moment decrease occurs in the same pressure domain. Nevertheless the lack of reproducibility of high pressure conditions does not allow a precise correlation between the structural and the magnetic transition when measured separately. A recent study [21] reports measurements of phonon dispersion curves by inelastic neutron scattering of bcc Iron under pressure up to 10 GPa over the entire stability range of the α phase. The authors find a surprisingly uniform pressure dependence of the α phase and a lack of any significant pretransitional change close to the bcc-hcp transition concluding that this behaviour confirms the driving role of magnetism of Iron into the transition.

In this Letter we report a simultaneous x-ray absorption spectroscopy (XAS) and x-ray magnetic circular dichroism (XMCD) study of the Fe bcc-hcp transition. The high sensitivity of XMCD allows us to follow the magnetic transition very precisely, and to correlate it to the local structure followed by XAS. We find a sharper transition domain for both the structural and the magnetic phase transition and a small pressure gap between them. Our results highlights the close relationship between the structural transition and the decrease of the magnetic moment.

Polarized x-ray absorption spectroscopy contains intrinsic structural, electronic and magnetic probes. These are respectively extended x-ray absorption fine structure (EXAFS), x-ray absorption near edge structure (XANES) and X-ray Magnetic Circular Dichroism (XMCD). The first two are able to differentiate clearly the signature of bcc or hcp local symmetry, while the third is very sensitive to polarized magnetic moments variation. Thus, for each pressure point, we can obtain simultaneous information on both the magnetic and structural properties of the system i.e. on the same sample and in the same thermodynamic conditions. As a consequence, at each measurement, there is no relative pressure incertitude. This is very important in the high-pressure domain where reproducibility of high-pressure hydrostatic conditions is difficult to obtain.

XMCD and XAS were recorded at the ESRF at the energy dispersive XAS beamline ID24 [22]. This station is designed to fulfill the requirements of detecting very small XMCD signals (down to ~ $10^{-3}$) under pressure at the K edges of 3d transition metals [23]. The sample (Goodfellow high purity 4 μm thick iron foil) was inserted into the Cu-Be Diamond Anvil Cell (DAC) and placed in a 0.4 T magnetic field parallel to the x-ray beam. Silicon oil was used as pressure transmitting medium and the pressure was measured with the ruby fluorescence technique [24]. Inherent to this pressure calibration technique, the absolute pressure incertitude is about 1 GPa, but the relative pressure incertitude between different pressure data points in the same run is 0.1 GPa. In this experiment, each XMCD spectrum was obtained by accumulating 200 XAS spectra with a change of direction of the applied magnetic field between two successive ones. By comparing the first and the last XAS spectra relative to one XMCD acquisition, it was possible to check the structural stability of the sample during the measurement. XMCD spectra were recorded from room pressure up to 22.4 ± 0.1 GPa. Several runs were carried out on freshly loaded samples, to check reproducibility.

Figure 1 compares the normalized Fe K-edge XAS (left panel) and XMCD (right panel) data of pure Fe foil at ambient conditions (bottom curves), to examples of spectra recorded within the DAC at different pressure values, between 10.0 ± 0.1 GPa and 22.4 ± 0.1 GPa, during a pressure increase ramp.

The energy range diffracted by the polychromator is limited to the XANES region and the first oscillations of the EXAFS domain. The XAS data shows very clearly that both the electronic structure and the local structure around Fe are drastically modified above 14 GPa: the pre-peak at 7116 eV becomes more resolved, the maximum of the absorption drifts towards higher energies (from 7132 to 7134 eV) and becomes less intense, and the frequency of the EXAFS oscillations is drastically reduced. The EXAFS signature of the different steps of the Fe bcc-hcp transition was already clearly identified by Wang and Ingalls [6] and our data are consistent with the structure reported.

The XMCD curves in panel b correspond to the simultaneously obtained XAS spectra shown in continuous lines in panel a. The amplitude of the ambient pressure XMCD signal obtained with the sample in the DAC (not shown) is equal to approximately 30 % of its value measured outside the DAC due to the smaller applied magnetic field which cannot overcome the demagnetization factor of the probed iron foil.

In order to better identify the onset and the evolution with pressure of this phase transition, each pair of XAS and XMCD spectra are carefully analyzed as following :

We performed the first derivative of the XAS spectra in the pressure range close to the transition region. The amplitude of the derivative signal changes drastically between the bcc phase and the hcp phase at E = 7137 eV, E =

7205 eV and E = 7220 eV. At these energy points, we have the highest sensitivity to the bcc/hcp phase fraction (note that the phase fraction calculated at the three different energies is identical within the error bar).
To quantify the amplitude of the XMCD signals, the absolute value of the background subtracted data are integrated in the energy range 7100 – 7122 eV. The magnetic/non-magnetic phase fraction is then determined.

In Figure 2 we plot (full squares) the evolution of the bcc/hcp phase fraction and of the magnetic/non-magnetic phase fraction (full circles) as a function of pressure. The onset of the structural transition occurs at about 13.8 ± 0.1 GPa and is over at 16.2 ± 0.1 GPa. In Ref. [6], the transition started at a lower pressure and occurred in a larger pressure range with respect to that observed in the present work. This could be attributed to the different conditions of non-hydrostaticity within the cell, and underlines the difficulty of reproducing the same thermodynamic conditions in different experiments. The magnetic transition is indeed quite abrupt occurring within a pressure range of 2.2 ± 0.2 GPa, with an onset at 13.5 ± 0.1 GPa. Our data shows that between 13.5 and 15.7 ± 0.1 GPa the XMCD is suddenly reduced to zero, within the error bars, indicating the disappearing of the room temperature ferromagnetic order of iron. The abrupt drop to zero of the iron magnetic moment when the bcc to hcp phase transition occurs proves the first order nature of the pressure induced magnetic transition, as predicted by Ekman et al.[10].

The narrower pressure transition domain for both the structural and the magnetic phase transition found in the present work with respect to previous structural and magnetic studies [6,19,20] could be attributed to a lower pressure gradient within the probed region, directly correlated to the smaller sample volume probed by the X-ray beam [22] with respect to previous experiments. The influence of different experimental conditions to the pressure values of transition onsets and to the pressure range of phase coexistence is well known and frequently addressed in the literature (see for example Ref. 6). For this reason, the simultaneous measurement of element specific magnetic properties using XMCD and of element specific local structural and electronic properties using XAS, that overcomes all uncertainty related to different hydrostatic conditions within the pressure cell and different probed volumes, is fundamental to address issues concerning correlation between magnetic, electronic and structural degrees of freedom of a system during phase transitions.

A closer look of the data obtained in the transition region shows that the value of the magnetic/non-magnetic phase fraction is systematically lower than of the bcc/hcp phase fraction for each XAS/XMCD pair of measurements during the phase transition. Thanks to the combination of the two probes, there is no relative pressure incertitude between XAS/XMCD data points. As a consequence it implies that our data shows that the drop of the magnetic moment occurs at slightly lower pressures with respect to the local structural modifications. This is better seen on the zoom of the transition region in figure 2.

Our observations are in good agreement with the transition pattern given in Ref. 10 where theoretical calculations show that the bcc iron phase instability is not primarily due to phonon softening but to the effect of pressure on the magnetism of Iron. In this scenario, the effect of pressure is essentially to enhance the overlap of the atomic wave functions, which broadens the flat d-bands and decreases the density of states at the Fermi level $D(E_f)$ below the stability limit of the ferromagnetism given by the Stoner criterion. The bcc phase becomes thermodynamically unstable with respect to the hcp phase. The scenario of a magnetic driven transition is also confirmed by other experimental work on the phonon dispersion of bcc Iron up to 10 GPa [21].

Ref. 10 predicts that the bcc phase retains very stable high-spin states with magnetic moments of approximately 2.0-2.2 $\mu_B$ even at very high pressure. It also predicts, for a specific transition path between the bcc and the hcp phase, the existence of a stable low-spin state with an intermediate magnetic moment of about 1 $\mu_B$ and an intermediate structure between bcc and hcp. We can also note that in a previous EXAFS study of the α−ε transition [6], an intermediate distorted bcc phase with an anomalously large lattice constant has been observed during the transition. Ref. [6] also reports the possibility of an intermediate fcc structure. Our study suggests that the iron high spin ferromagnetic state disappears before the complete transformation into hcp. This could also be due to the presence of magnetically dead layers at the interface between the bcc and hcp phase or to the low spin state with an order temperature smaller than room temperature. The importance in the bcc to hcp phase transformation of the shear stresses highlighted by ref. [14] could be also investigated. Further studies on the transition hysteresis and at low temperature should provide fundamental information for a better understanding of the transition dynamics.

The observed absence of macroscopic magnetization above 15.7 GPa is in good agreement with the recent prediction of anti-ferromagnetic fluctuations for a small pressure range starting with the onset of hcp phase [15], which is suggested as a possible origin of the superconductivity.

We have measured XAS and XMCD of iron metal under pressure along the bcc to hcp phase transition. The simultaneous measurement of element specific magnetic properties using XMCD and of element specific local structural and electronic properties using XAS has an enormous potential in correlating the magnetic, electronic and structural degrees of freedom during phase transitions. For Fe at high pressure, we find that the local structure and the magnetic transition occur within 2.4 ± 0.2 GPa and are much sharper than usually described in literature. This proves unambiguously the first order nature of the iron bcc to hcp transition. We have also evidence a small pressure gap between magnetic transition and the structural one suggesting the driving role of magnetism during the transition.

References

[1] H. Hasegawa and D.G. Pettifor Phys. Rev. Lett. **50**, 130 (1983).

[2] L. Stixrude, R.E. Cohen and D.J. Singh, Phys. Rev. B **50,** 6442 (1994).

[3] R. Jeanloz, Ann. Rev. Earth Planet, Sci **18**, 357 (1990).

[4] D. Bancroft, E. L. Peterson and S. Minshall, J. Appl. Phys. **27**, 291 (1956).

[5] M. Nicol and G. Jura, Science **141**, 1035 (1963).

[6] F.M. Wang and R. Ingalls Phys. Rev. B **57**, 5647 (1998).

[7] K. Shimizu, T. Kimura, S. Furomoto, K. Takeda, K. Kontani, Y. Onuki, K. Amaya, Nature **412**(6844) 316 (2001).

[8] S.K. Bose, O.V. Dolgov, J.Kortus, O.Jepsen, and O.K. Andersen Phys. Rev. B **67**, 214518 (2003).

[9] T. Asada and K. Terakura, Phys. Rev. B **46**, 13599 (1992).

[10] M. Ekman, B. Sadigh, K. Einarsdotter and P. Blaha, Phys. Rev. B **58**, 5296 (1998).

[11] H. C. Herper, E. Hoffmann and P. Entel, Phys. Rev. B **60**, 3839 (1999).

[12] P. Söderlind and J. A. Moriarty, Phys. Rev. B **53**, 14063 (1996).

[13] W. Pepperhoff and M. Acet, Constitution and magnetism of Iron and its alloys (Springer-Verlag, Berlin Heidelberg, 2001).

[14] K. J. Caspersen, A. Lew, M. Ortiz and E. A. Carter, *Phys. Rev. Lett*. **93**, 115501 (2004).

[15] V.Thakor, J.B. Stauton,  J. Poulter, S. Ostanin, B. Ginatempo and E. Bruno, Phys. Rev. B **67**, 180405 (R) (2003).

[16] E. Huang, W. A. Basset and P. Tao, J. Geophys Res. **92**, 8129 (1987).

[17] G. Cort, R. D. Taylor and J. 0. Willis, J. Appl. Phys. **53**,2064 (1982).

[18] R D. Taylor, G. Cort and J. 0. Willis, J. Appl. Phys. **53**, 8199 (1982)

[19] R.D. Taylor, M.P. Pasternak and R. Jeanloz, J. Appl. Phys. **69**, 6126 (1991).

[20] J.P. Rueff, M. Krisch, Y. Q. Cai, A. Kaprolat, M. Hanfland, M. Lorenzen, C. Maschiovecchio, R. Verbeni and F. Sette, Phys. Rev. B **60**, 14510, (1999).

[21] S. Klotz and M. Braden, Phys. Rev. Lett. **85**, 3209 (2000).

[22] S. Pascarelli, O. Mathon and G. Aquilanti, J. of Alloys and Compounds **362**, 33 (2004).

[23] O. Mathon, F. Baudelet, J.-P. Itié, S. Pasternak, A. Polian and S. Pascarelli, J. Synchr. Rad.**11**, 423 (2004).

[24] J.C. Chervin, B. Canny and M. Mancinelli, High Pressure Research **21**, 305 (2001).

Figure captions

Fig. 1. Fe K-edge XAS (left panel) and some example of XMCD (right panel) as a function of pressure between the ambient pressure bcc phase and the high pressure hcp phase. Continuous lines correspond to examples of data measured simultaneously. The small glitches in the XAS at 7158 and 7171 eV in the high pressure data are artifacts due to small defects of the polychromator crystal.

Fig. 2. Evolution of the amplitude of the derivative of the XAS (full squares) compared to the reduction of the amplitude of the XMCD signals (full circles). The inset is a zoom of the transition region, the two lines are a linear fit of the XAS and XMCD data points in this pressure region.

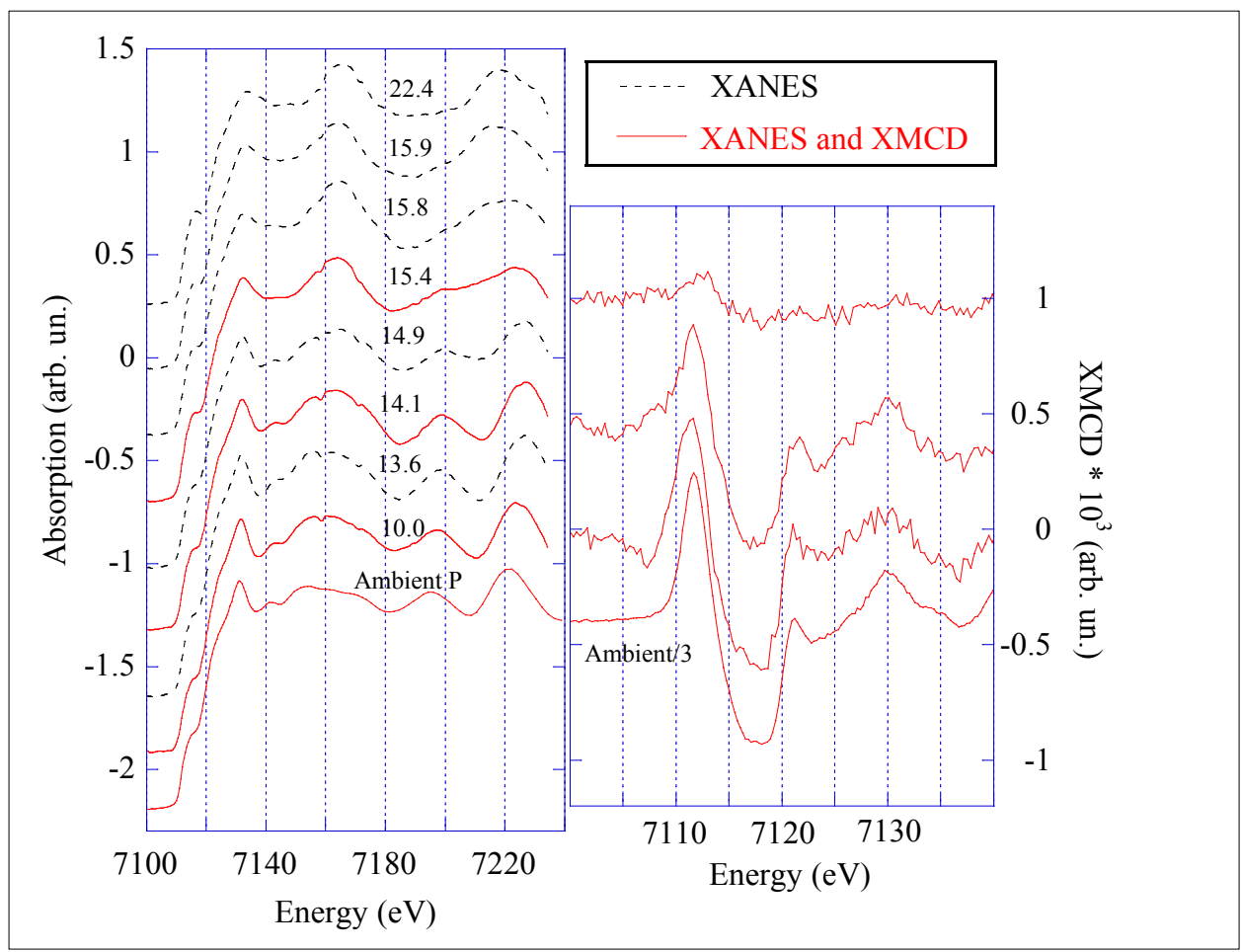

Fig. 1.

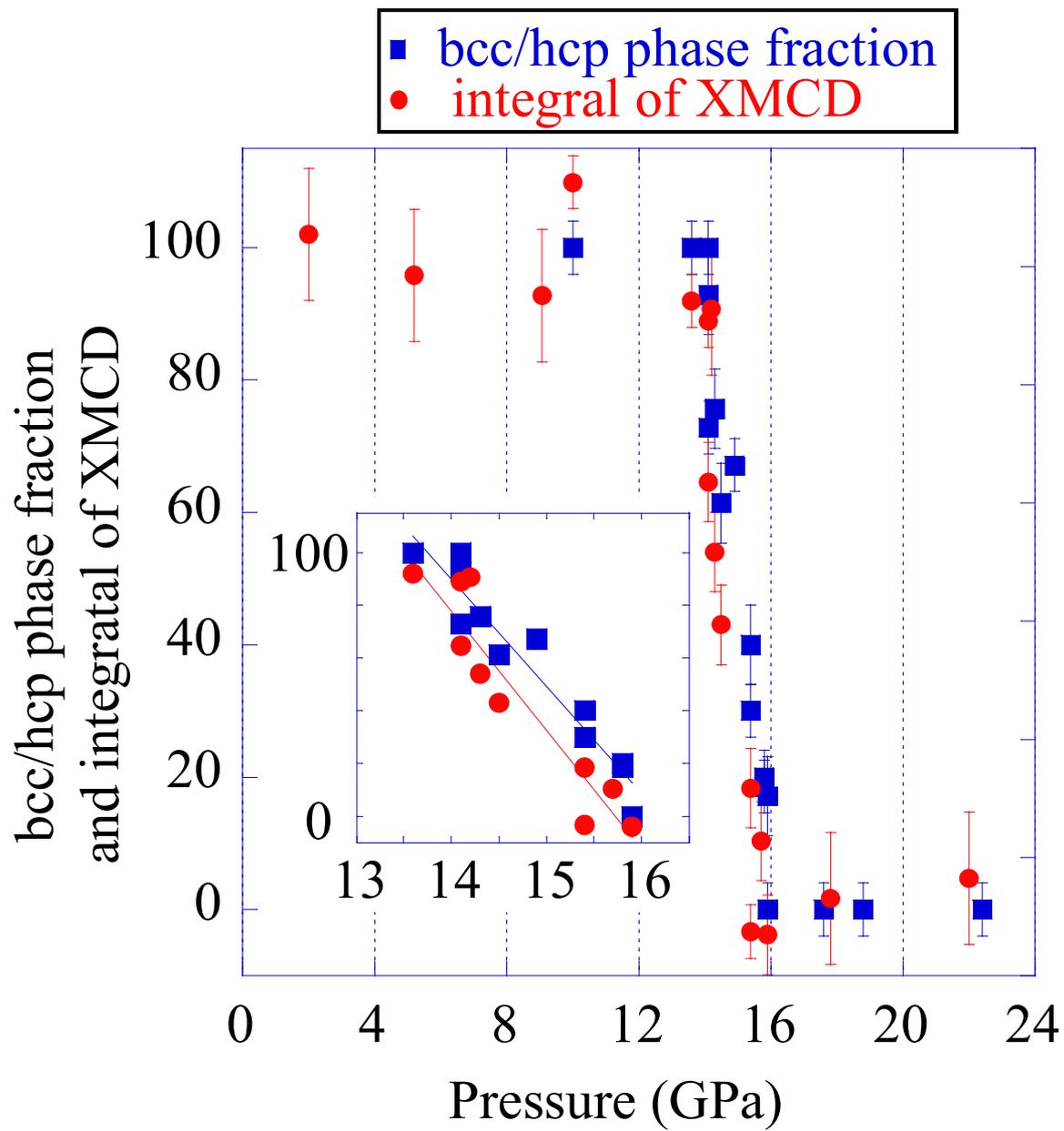

Fig. 2.